# Linked by Loops:
## Network Structure and Switch Integration in Complex Dynamical Systems

Dennis Cates Wylie*

Department of Environmental Science, Policy, and Management, University of California, Berkeley, dennisw@socrates.berkeley.edu

**Abstract**
Simple nonlinear dynamical systems with multiple stable stationary states are often taken as models for switchlike biological systems. This paper considers the interaction of multiple such simple multistable systems when they are embedded together into a larger dynamical "supersystem." Attention is focused on the *network structure* of the resulting set of coupled differential equations, and the consequences of this structure on the propensity of the embedded switches to act independently versus cooperatively. Specifically, it is argued that both larger average and larger variance of the node degree distribution lead to increased *switch independence*. Given the frequency of empirical observations of high variance degree distributions (e.g., power-law) in biological networks, it is suggested that the results presented here may aid in identifying switch-integrating subnetworks as comparatively homogenous, low-degree, substructures. Potential applications to ecological problems such as the relationship of stability and complexity are also briefly discussed.

**PACS**: 89.75.Hc, 89.75.Fb, 87.17.Aa, 87.23.Cc

**Keywords**: systems biology, multistability, complex systems, feedback loops, dynamics on networks

## 1. Introduction

Many biological systems contain various subsystems which exhibit switch-like behavior [1]: while stable to suitably small perturbations of their conditions, they may be observed to jump suddenly to a new state in response to sufficient provocation. This behavior arises naturally in nonlinear dynamical models with multiple stable fixed points, and it is thus not surprising that these models are frequently invoked in the study of such subsystems [1, 2].

Assuming that this approach is successful in capturing the essential features of this or that individual switch, one might next step back and ask: how, then, do the switches work when reassembled together into the larger biological context from which they were originally wrested? In other words, what happens to that switch there when I toggle this one here?

To such a general question there can be only one sensible answer: it depends. But it may be hoped that at least some of the factors on which it depends are structural features not entirely remote from our observation. For many biological systems, the most accessible data available comes in the form of network structure [3-10]. Thus emerges the topic of this paper: how does the integration of multiple switches into a common "supersystem" depend on network structure?

It is perhaps wise at this point to pause and consider some specific biological contexts in which switch integration might be expected to be an essential feature. The processes of cellular determination and differentiation would appear natural candidates. Switchlike multistability has long been thought to be an important feature in differentiation [11-14], and the feedback-loop-linked modular structures of the genetic regulatory networks underlying development [15, 16] suggest linked local switches. Likewise, decision-making by a modular nervous system [17] seems a tempting target for this approach.

*Corresponding Author: Dennis Cates Wylie
dennisw@berkeley.edu

However, the likely field of most immediate consideration for switch integration modeling is community ecology [18]. Applications of the theory of nonlinear dynamics have long been common in ecology, and the concepts of keystone species and indirect effects [18, 19] bring up questions regarding the propagation of local (i.e., one or a few node) perturbations through the network of species making up a community. Viewed through the lens of network theory, these questions share similarities with the problem of switch integration defined here. The long-running stability-complexity debate [20-23] may also overlap with the ideas presented in this paper; this in particular is further pursued in **section 5**.

The analysis of nonlinear dynamical systems in terms of their network structure is an old and established field [12, 21, 24-30], with the potential for many new developments given the current enthusiasm for and rapid development of network theory [31-37]. Consider, for example, the recent work studying the phenomenon of dynamic synchronization on network structures [38-40], which may prove interesting to compare with the problem of switch integration.

The techniques employed in this paper, however, owe a special debt to the work of Richard Levins [41]. Building on basic ideas from the study of the stability of control systems [42, 43], Levins illuminated a connection between the characteristic polynomial and the feedback loops of a sparse matrix. This connection provides the basis for the techniques used herein, as described in **section 3**.

**2. Overview of Switch Integration**

*Switch independence* is here defined by the following system property: if a switch setting is (stably) available for one available combination of the settings of any other switches present, it must be available for *all* available combinations of the settings of the other switches (see **figure 1**). A precise statement of what is meant by the term "switch setting," at least with regard to the particular model systems considered herein, is offered in **appendix 1**. A system of switches co-embedded in a network may then be said to have the property of *switch integration* inasmuch as it lacks that of switch independence.

Results regarding the effects of network topology on switch integration were arrived at through a combination of computer simulation and theoretical argument. The details of the computer simulations may be found in **appendix 1**. In each such simulation, two bistable subsystems were embedded together into a larger randomly generated network (in which nodes represent state variables and arcs represent the pattern of dynamic interactions (see **section 3**)) constructed as described in **appendices 1-2**. Note that, prior to coupling to the system network variables, there must then be four distinct stable fixed points (described in terms of switch settings as (off,off), (off,on), (on,off), and (on,on)) available to the two bistable switches considered as a single system. Those systems in which exactly two of these four fixed points remained stable after coupling of the switches to the network were examined for the property of switch integration (with results for different network types shown in **figure 2**). The various types of reactions and associated parameters in the model systems described in **appendix 1** were chosen on the grounds of simplicity (e.g., all dynamics quadratic functions of the state variables) and (bi)stability (i.e., in such a way that a large fraction of the systems constructed exhibited two stable fixed points).

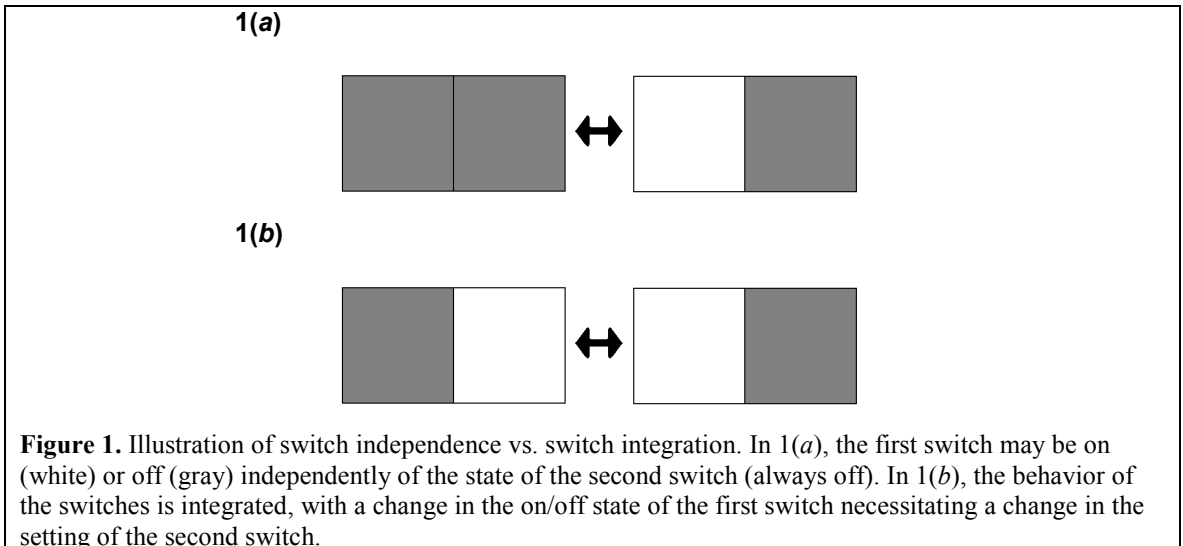

**Figure 1.** Illustration of switch independence vs. switch integration. In 1(*a*), the first switch may be on (white) or off (gray) independently of the state of the second switch (always off). In 1(*b*), the behavior of the switches is integrated, with a change in the on/off state of the first switch necessitating a change in the setting of the second switch.

Perhaps the most immediately apparent trend in **figure 2** is that the propensity for switch integration declines rapidly with increasing system dimensionality, regardless of which sort of network structure is considered. This result is likely explained by the observation that the shortest path length through the network connecting the two switches (in either direction) generally increases as the number of nodes in the system increases, leading to weaker coupling between the switches. Note that this indicates that, for switch integration in larger systems, it is probably necessary for the switches in question to be positioned nearer to each other than they would be if they were placed randomly, as done here.

One might then suppose that, on the basis of switch-distance considerations alone, increasing the arc density of a network should lead to increased switch integration by increasing the intensity of interactions between the switches. However, upon consulting **figure 2(*a*)**, one sees the opposite trend with regard to network density: the denser networks tend toward *reduced* switch integration.

Examining **figure 2(*b*)**, it is apparent that increasing the variance in the node degree distribution (as occurs for scale-free digraphs of increasing size – see **appendix 2**) also tends to reduce the likelihood of switch integration. Note that higher variance of degree distribution tends to imply that the neighbors of a randomly selected node will tend to have increasingly greater than average degree, since vertices with many connections ("hubs") neighbor many more nodes than do vertices of low degree. (This effect also underlies the explosive dynamics of epidemic spreading processes on high degree-variance network topologies, and is discussed in more detail in that context in Meyers [44].)

Thus, one might hypothesize that the effects of both network density and degree distribution variance might be understood in terms of the relationship between switch integration and local connection density in the neighborhood(s) of the switches to be integrated. **Section 4** develops the *differential overlap dependence hypothesis* (DODH) as a theory of this relationship and presents further simulation results supporting the DODH (see also **appendices 3-4**).

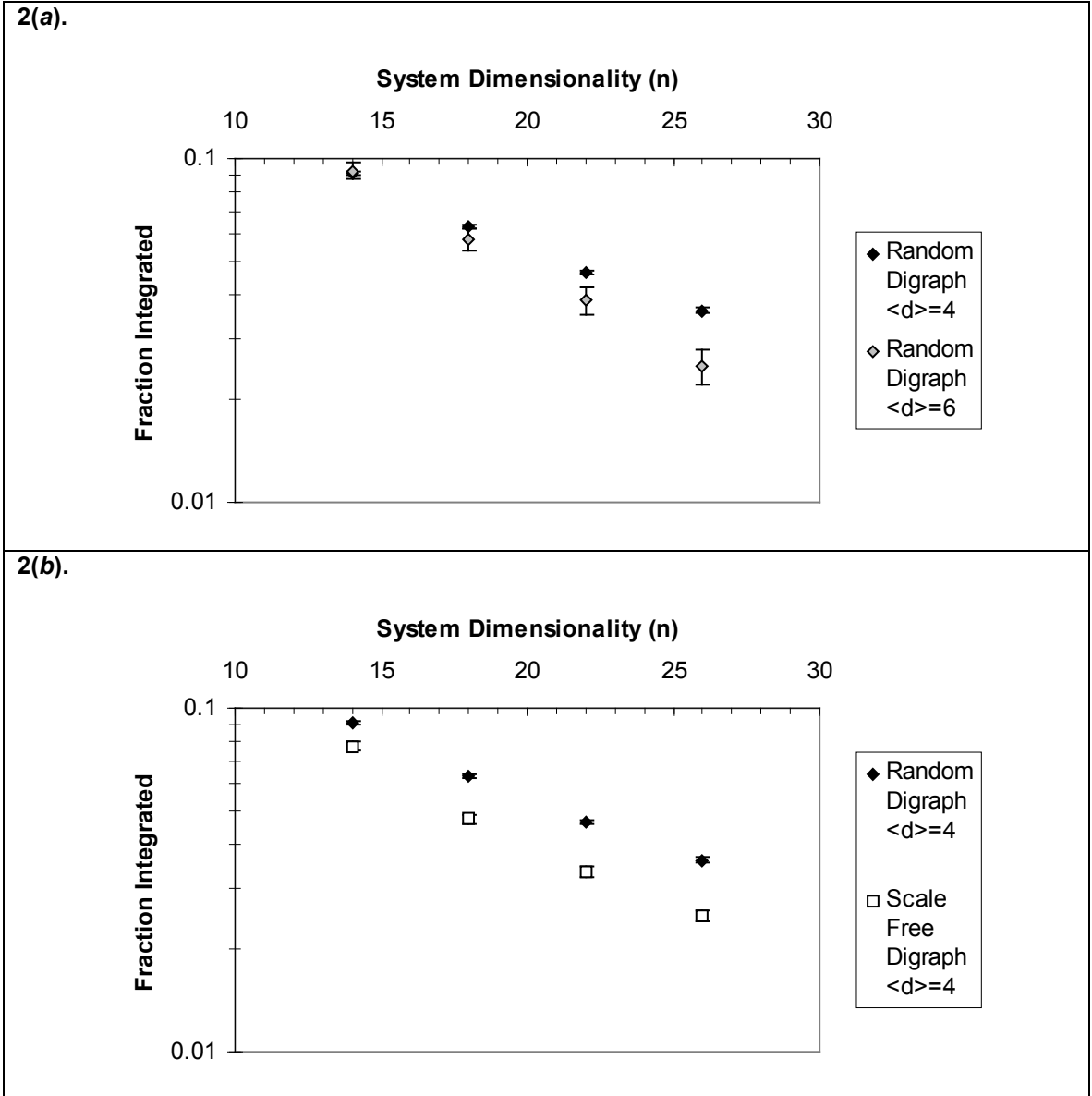

**Figure 2.** Fraction of systems built and satisfying criteria for consideration as described in **appendix 1** which exhibited switch integration (as defined in **appendix 1**). Digraph topologies are constructed as in **appendix 2** (<*d*>=average degree). Error bars indicate standard error in estimation of fraction integrated.

The DODH is built upon the relationship of the characteristic polynomial $F_k(\mathbf{x}^\alpha)$ of a system at a fixed point $\mathbf{x}^\alpha$ and the system's network topology (see **section 3**). As the stability of a fixed point $\mathbf{x}^\alpha$ is determined by the values of the quantities $F_k(\mathbf{x}^\alpha)$ (by, e.g., the Routh-Hurwitz criterion [42]), it is here suggested that the the correlation

(1) $$\text{Corr}(F_k(\mathbf{x}^\alpha), F_k(\mathbf{x}^\beta)) = \frac{\langle\langle F_k(\mathbf{x}^\alpha) F_k(\mathbf{x}^\beta) \rangle\rangle}{\sqrt{\langle\langle F_k(\mathbf{x}^\alpha) F_k(\mathbf{x}^\alpha) \rangle\rangle \langle\langle F_k(\mathbf{x}^\beta) F_k(\mathbf{x}^\beta) \rangle\rangle}}$$

(where the fixed points $\mathbf{x}^\alpha$ and $\mathbf{x}^\beta$ correspond to different settings of a single switch) under stochastic system perturbation may be predictive of the correlation of the stabilities of $\mathbf{x}^\alpha$ and $\mathbf{x}^\beta$ under such perturbation. (Note that any perturbation which changes the phase space location of the fixed points $\mathbf{x}^\alpha$ and $\mathbf{x}^\beta$ will generally result in changes to the values of the corresponding $F_k(\mathbf{x}^\alpha)$ and $F_k(\mathbf{x}^\beta)$ as well.)

According the DODH, $Corr(F_k(\mathbf{x}^\alpha),F_k(\mathbf{x}^\beta))$ should be expected to increase with both the average and variance of the node degree distribution of the system network (see **section 4**). Computational results (on systems perturbed by random modification of the **x**-independent constant terms in the dynamics $d\mathbf{x}/dt=\mathbf{f}(\mathbf{x})$) for comparison with this prediction are presented in **section 4** (and discussed in more detail in **appendix 3**).

Consider now the effect of changing the setting of a switch $S$ on the various settings of another switch $S'$. The phase space positions of the fixed points representing the settings of $S'$ will all be modified as a result of the shift in setting of $S$. That is, this shift in $S$-setting may be regarded as a perturbation to the settings of $S'$. Applying the DODH predictions regarding the correlation of fixed point stabilities to this situation, it would thus be expected that in networks with higher arc density or degree-distribution variance, the likelihood of selectively destabilizing only one (as opposed to both or neither) of the two settings of $S'$ would be less than it would be for sparser or more homogenous networks. Noting that "switch integration" as defined here is exactly this phenomenon of selective destabilization of some (but not all) settings of $S'$ as a result of the shifting of the setting of another switch $S$, the DODH predicts that higher average or variance of degree distribution decreases the likelihood of switch integration, in agreement with the results of **figure 2**.

**3. Relationship Between Network Structure and Linearized Dynamics of a System**
*3.1 Graphical Interpretation of the Characteristic Polynomial of a Matrix*

As discussed in Puccia and Levins [41], the characteristic polynomial of a matrix (taken here to be the matrix

(2) $$M_{ij}(\mathbf{x}^\alpha) = \frac{\partial f_j}{\partial x_i}(\mathbf{x}^\alpha)$$

representing the linearized dynamics of a system $d\mathbf{x}/dt=\mathbf{f}(\mathbf{x})$ at the fixed point $\mathbf{x}^\alpha$) may be written (using $I$ to represent the identity matrix)

(3) $$\mathrm{Det}(M(\mathbf{x}^\alpha) - \lambda I) = (-1)^{n-1} \sum_{k=0}^{n} F_{n-k}(\mathbf{x}^\alpha) \lambda^k$$

where the coefficients $F_{n-k}(\mathbf{x}^\alpha)$, defined precisely in **equation (4)** below, are sums of terms which correspond to loop structures in an associated digraph.

For the purposes of this paper, a digraph [45] is taken to be a set of nodes along with a set of directed arcs, each of which begins at one node (its "tail") and ends at another (its "head"). It is not allowed for two distinct arcs to have both the same tail and same head in the same digraph. It is, however, here allowed for one arc to have the same node as its tail and its head.

A digraph may be associated with a given $n$-dimensional matrix $M$ (see **figure 3**): first assign $n$ numbered nodes $\{1,2,3,\ldots,n\}$, then add an arc $(i \rightarrow j)$ iff $M_{ij} \neq 0$. This is generalized to a rule for associating digraphs with dynamical systems $d\mathbf{x}/dt=\mathbf{f}(\mathbf{x})$ by letting $M_{ij}=\partial_i f_j$ (**equation (2)**). Of course, the question arises as to where in phase space to evaluate the matrix $\partial_i f_j$; it proves convenient to adopt the convention that the associated digraph has arc $(i \to j)$ present iff there exists **x** somewhere in phase space such that $\partial_i f_j(\mathbf{x}) \neq 0$. It is apparent from this definition that the digraph associated with a dynamical system does not depend on choice of a specific fixed point, i.e., that system topology is the same for all fixed points.

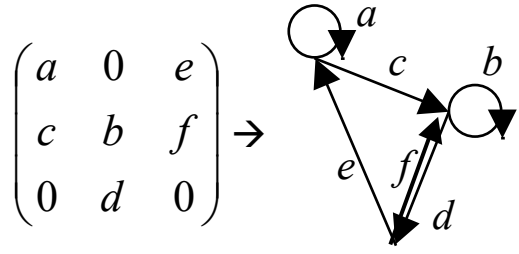

**Figure 3.** Weighted digraph associated with a matrix, as described in **section 3.1**. There are two 1-terms in this system (the one-loops with weights $a$ and $b$), two 2-terms (the set of both one-loops, with total weight –$(a)(b)$ (negative b/c made up of two loops), and the two-loop with weight $df$), and two 3-terms, with weights –$(a)(df)$ and $cde$. Thus, the coefficients of the characteristic polynomial of this matrix are $F_1=(a+b)$, $F_2=(-ab+df)$, and $F_3=(-adf+cde)$ (along with the trivial $F_0=-1$).

A "path" from node $i$ to node $j$ in a digraph will be here defined as an ordered set $P$ of arcs present in the digraph such that each arc has as its tail the head of the previous arc and as its head the tail of the next arc. The first arc in $P$ has as its tail node $i$, while the last arc in $P$ has as its head node $j$. A "loop" in a digraph is then defined as a path in which the starting node $i$ coincides with the ending node $j$ (so that $i=j$; more precisely, a loop is any such set of arcs, forgetting the arbitrary choice of base node $i$). Note that loops of length one are here allowed.

It is useful also to give a name to collections of disjoint (i.e., not sharing any nodes) loops which pass through exactly $k$ nodes (and hence have exactly $k$ arcs in total); such structures will be called "$k$-terms." This name is chosen because it turns out that the terms present in the sum $F_k$ for a particular matrix are in bijective correspondence with the $k$-terms present in the associated digraph [41]. More specifically, each $k$-term structure $K$ present in the associated digraph of a matrix contributes the product of the matrix entries associated with the arcs of $K$ to the coefficient $F_k$ of the characteristic polynomial (with an additional sign factor depending on the number of disjoint loops $c(K)$ composing $K$). That is,

$$(4) \quad F_k(\mathbf{x}^\alpha) = \sum_{K \in \Theta_k} (-1)^{c(K)+1} \left[ \prod_{(i \to j) \in K} M_{ij}(\mathbf{x}^\alpha) \right]$$

where the sum runs over the set $\Theta_k$ of all possible distinct $k$-terms $K$ (two $k$-terms are distinct as long as they do not contain the same arc set, ignoring ordering), and the product runs over all arcs $(i \to j)$ contained within $K$. An example of the application of **equation (4)** to a particular matrix is offered in **figure 3** and its caption. It is convenient to introduce the notation $K(\mathbf{x}^\alpha)$ to denote the numerical value associated with the $k$-term $K$ in **equation (4)** above, i.e.,

$$(5) \quad K(\mathbf{x}^\alpha) = (-1)^{c(K)+1} \prod_{(i \to j) \in K} M_{ij}(\mathbf{x}^\alpha)$$

so that **equation (4)** may be rewritten as

(6) $$F_k(\mathbf{x}^\alpha) = \sum_{K \in \Theta_k} K(\mathbf{x}^\alpha)$$

The notation $F_k$ (suggested by Levins [41]) for the coefficients of the characteristic polynomial is intended to suggest "feedback at level $k$." The content of **equation (6)** is then that the $k$-feedback of a system (at a particular fixed point $\mathbf{x}^\alpha$) is essentially a weighted sum of all the $k$-terms present in the system's topology, with the weightings arising from the linearized dynamics. Considering the disjoint loops composing an arbitrary $k$-term as "feedback loops," the idea underlying the interpretation of $F_k$ as $k$-feedback is laid bare.

Note that the matrix entries $M_{ij}(\mathbf{x}^\alpha)$ are signed quantities: if the product of all of these arc weightings for the arcs present in a particular loop is positive, the loop in question may be called a positive feedback loop; negative feedback loops are defined analogously. The sign factor $(-1)^{c(K)+1}$ appearing in **equation (5)** may then be understood as necessary to ensure that the overall contribution to $k$-feedback $F_k$ of a $k$-term $K$ containing $c(K)$ all-negative disjoint feedback loops is negative: that is,
$$(-1)^{c(K)+1}(-1)^{c(K)} = -1$$
More generally, a $k$-term $K$ will provide a negative contribution to $k$-feedback $F_k$ iff an even number of the disjoint feedback loops composing it are positive (with the remainder negative). A necessary, but not sufficient, condition for the stability of a fixed point is that total $k$-feedback $F_k$ must be negative for all $k$ [41].

**Equations (3)-(4)** may be derived by considering the bijection between permutations (in terms of which determinants are usually defined) and $k$-terms which may be seen in the common cycle notation for permutations [46]. For example, the permutation (12345)(678) corresponds to the $k$-term

(7) $K_{(12345)(678)} = \{(1\rightarrow 2), (2\rightarrow 3), (3\rightarrow 4), (4\rightarrow 5), (5\rightarrow 1), (6\rightarrow 7), (7\rightarrow 8), (8\rightarrow 6)\}$

*3.2 Graphical Interpretation of Characteristic Polynomial Covariance*

If $\mathbf{x}^\alpha$ and $\mathbf{x}^\beta$ are both fixed points of the dynamical system $d\mathbf{x}/dt = \mathbf{f}(\mathbf{x})$, it is straightforward to see that the product $F_k(\mathbf{x}^\alpha)F_l(\mathbf{x}^\beta)$ will admit a topological interpretation as well. Specifically, $F_k(\mathbf{x}^\alpha)F_l(\mathbf{x}^\beta)$ will decompose as a sum over all possible combinations of $k$-terms (with arc weights taken from the matrix of the linearization at $\mathbf{x}^\alpha$) and $l$-terms (with arc weights from $\mathbf{x}^\beta$),

(8) $$F_k(\mathbf{x}^\alpha)F_l(\mathbf{x}^\beta) = \left(\sum_{K \in \Theta_k} K(\mathbf{x}^\alpha)\right)\left(\sum_{L \in \Theta_l} L(\mathbf{x}^\beta)\right)$$
$$= \sum_{\{K,L\} \in \Theta_k \times \Theta_l} [K(\mathbf{x}^\alpha)L(\mathbf{x}^\beta)]$$

Any particular such combination of a $k$-term $K$ and an $l$-term $L$ defines a graphical structure $A$ of its own (the union of the two arc sets involved – see **figure 4**) – a "$(k,l)$-term." **Equation (8)** may then be rewritten as

(9) $$F_k(\mathbf{x}^\alpha)F_l(\mathbf{x}^\beta) = \sum_{A \in \Theta_{k,l}} \sum_{\{\{K_A,L_A\} \in \Theta_k \times \Theta_l | K_A \cup L_A = A\}} [K_A(\mathbf{x}^\alpha)L_A(\mathbf{x}^\beta)]$$

with the outer sum in **equation (9)** now running over the set $\Theta_{k,l}$ of all distinct "$(k,l)$-terms" $A$ (distinct again meaning that the arc sets in question are distinct ignoring ordering) while the inner sum runs over all distinct pairs $\{K_A, L_A\}$ of $k$-term $K_A$ and $l$-term

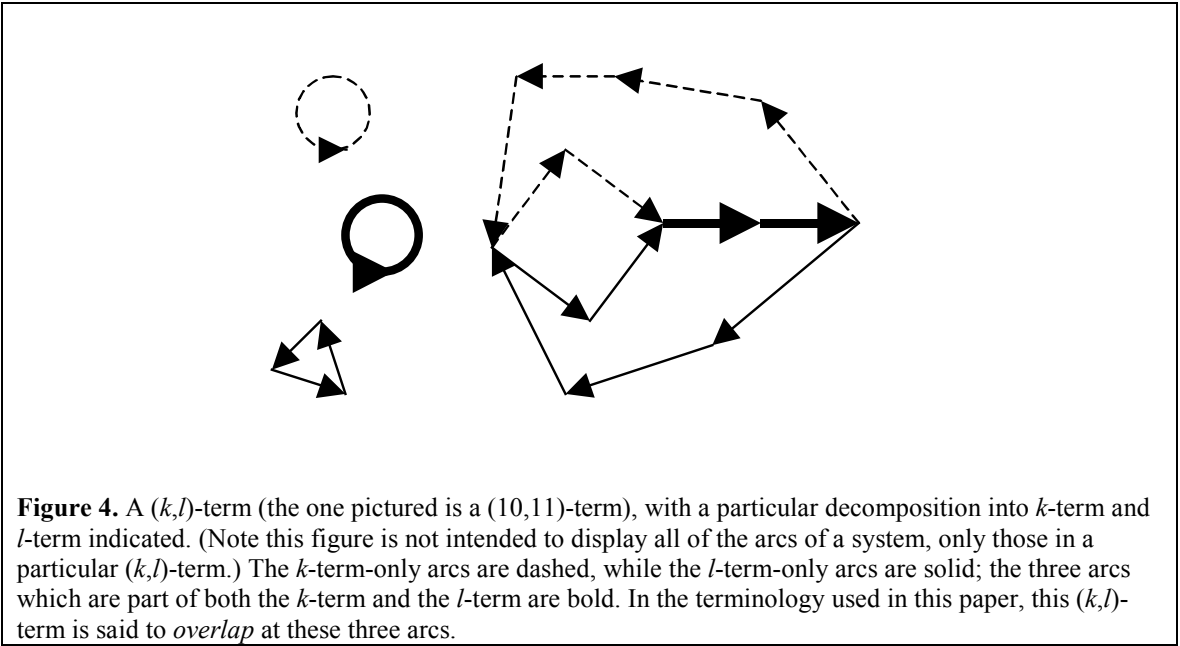

**Figure 4.** A ($k,l$)-term (the one pictured is a (10,11)-term), with a particular decomposition into $k$-term and $l$-term indicated. (Note this figure is not intended to display all of the arcs of a system, only those in a particular ($k,l$)-term.) The $k$-term-only arcs are dashed, while the $l$-term-only arcs are solid; the three arcs which are part of both the $k$-term and the $l$-term are bold. In the terminology used in this paper, this ($k,l$)-term is said to *overlap* at these three arcs.

$L_A$ whose union results in the particular ($k,l$)-term $A$. For notational convenience the set of such pairs is defined as $D^{(k,l)}{}_A$ (note that $k$ and $l$ must be specified, as there may also be ways of decomposing $A$ into $k_2$-term and $l_2$-term with $k_2 \neq k$ and $l_2 \neq l$); that is,

(10) $\quad D_A^{(k,l)} = \{\{K_A, L_A\} \in \Theta_k \times \Theta_l \mid K_A \cup L_A = A\}$

This notation is used below in writing summations like that of **equation (9)**.

If the arc weights $M_{ij}(\mathbf{x}^\alpha)$ which determine the $F_k(\mathbf{x}^\alpha)$ are probabilistically distributed quantities, the graphical interpretation of the product $F_k(\mathbf{x}^\alpha)F_l(\mathbf{x}^\beta)$ translates immediately into a similar expression for the moment $<F_k(\mathbf{x}^\alpha)F_l(\mathbf{x}^\beta)>$,

(11) $\quad \left\langle F_k(\mathbf{x}^\alpha) F_l(\mathbf{x}^\beta) \right\rangle = \sum_{A \in \Theta_{k,l}} \sum_{\{K_A, L_A\} \in D_A^{(k,l)}} \left\langle K_A(\mathbf{x}^\alpha) L_A(\mathbf{x}^\beta) \right\rangle$

In an entirely analogous manner, a graphical expression for $<F_k(\mathbf{x}^\alpha)><F_l(\mathbf{x}^\beta)>$ is obtained in terms of appropriate products of expectation values of products of arc weights,

(12) $\quad \left\langle F_k(\mathbf{x}^\alpha) \right\rangle \left\langle F_l(\mathbf{x}^\beta) \right\rangle = \sum_{A \in \Theta_{k,l}} \sum_{\{K_A, L_A\} \in D_A^{(k,l)}} \left\langle K_A(\mathbf{x}^\alpha) \right\rangle \left\langle L_A(\mathbf{x}^\beta) \right\rangle$

Thus, the graphical interpretation of $F_k(\mathbf{x}^\alpha)F_l(\mathbf{x}^\beta)$ extends to the covariance $<<F_k(\mathbf{x}^\alpha)F_l(\mathbf{x}^\beta)>> = <F_k(\mathbf{x}^\alpha)F_l(\mathbf{x}^\beta)> - <F_k(\mathbf{x}^\alpha)><F_l(\mathbf{x}^\beta)>$ through the formula

(13) $\quad \left\langle\!\left\langle F_k(\mathbf{x}^\alpha) F_l(\mathbf{x}^\beta) \right\rangle\!\right\rangle = \sum_{A \in \Theta_{k,l}} \sum_{\{K_A, L_A\} \in D_A^{(k,l)}} \left\langle\!\left\langle K_A(\mathbf{x}^\alpha) L_A(\mathbf{x}^\beta) \right\rangle\!\right\rangle$

## 4. The Differential Overlap Dependence Hypothesis

To develop the differential overlap-dependence hypothesis, first consider the fact that the stochastically perturbed weight $M_{ij}(\mathbf{x}^\alpha)$ of an arc ($i \rightarrow j$), like any probabilistically distributed quantity, is better correlated with itself than it is with any other quantity, including in particular the value of the same arc weighting at a different fixed point $\mathbf{x}^\beta$:

(14)
$$\frac{\left|\left\langle\left\langle M_{ij}(\mathbf{x}^\alpha)M_{ij}(\mathbf{x}^\beta)\right\rangle\right\rangle\right|}{\sqrt{\left|\left\langle\left\langle M_{ij}(\mathbf{x}^\alpha)^2\right\rangle\right\rangle\left\langle\left\langle M_{ij}(\mathbf{x}^\beta)^2\right\rangle\right\rangle\right|}} \leq 1 = \frac{\left|\left\langle\left\langle M_{ij}(\mathbf{x}^\alpha)M_{ij}(\mathbf{x}^\alpha)\right\rangle\right\rangle\right|}{\sqrt{\left|\left\langle\left\langle M_{ij}(\mathbf{x}^\alpha)^2\right\rangle\right\rangle\left\langle\left\langle M_{ij}(\mathbf{x}^\alpha)^2\right\rangle\right\rangle\right|}}$$

(i.e., $|\text{Corr}(M_{ij}(\mathbf{x}^\alpha), M_{ij}(\mathbf{x}^\beta))| \leq 1$). Now let $\Pi_1(\mathbf{x}^\alpha)$ be the product of the weightings of some set of arcs taken at the fixed point $\mathbf{x}^\alpha$ and define $\Pi_2(\mathbf{x}^\beta)$ similarly as the product of weightings of some different set of arcs taken at the fixed point $\mathbf{x}^\beta$. *If* the weights of all the various distinct arcs in the system were statistically independent of one other, then it would follow that (assuming $(i \to j)$ is not one of the arcs making up $\Pi_1$ or $\Pi_2$):

(15)
$$\frac{\left|\left\langle[M_{ij}(\mathbf{x}^\alpha)\Pi_1(\mathbf{x}^\alpha)][M_{ij}(\mathbf{x}^\beta)\Pi_2(\mathbf{x}^\beta)]\right\rangle\right|}{\sqrt{\left|\left\langle[M_{ij}(\mathbf{x}^\alpha)\Pi_1(\mathbf{x}^\alpha)][M_{ij}(\mathbf{x}^\alpha)\Pi_2(\mathbf{x}^\alpha)]\right\rangle\left\langle[M_{ij}(\mathbf{x}^\beta)\Pi_1(\mathbf{x}^\beta)][M_{ij}(\mathbf{x}^\beta)\Pi_2(\mathbf{x}^\beta)]\right\rangle\right|}}$$
$$= \frac{\left|\left\langle M_{ij}(\mathbf{x}^\alpha)M_{ij}(\mathbf{x}^\beta)\right\rangle\right|}{\sqrt{\left|\left\langle M_{ij}(\mathbf{x}^\alpha)^2\right\rangle\left\langle M_{ij}(\mathbf{x}^\beta)^2\right\rangle\right|}} \frac{\left|\left\langle\Pi_1(\mathbf{x}^\alpha)\Pi_2(\mathbf{x}^\beta)\right\rangle\right|}{\sqrt{\left|\left\langle\Pi_1(\mathbf{x}^\alpha)\Pi_2(\mathbf{x}^\alpha)\right\rangle\left\langle\Pi_1(\mathbf{x}^\beta)\Pi_2(\mathbf{x}^\beta)\right\rangle\right|}}$$
$$\leq \frac{\left|\left\langle\Pi_1(\mathbf{x}^\alpha)\Pi_2(\mathbf{x}^\beta)\right\rangle\right|}{\sqrt{\left|\left\langle\Pi_1(\mathbf{x}^\alpha)\Pi_2(\mathbf{x}^\alpha)\right\rangle\left\langle\Pi_1(\mathbf{x}^\beta)\Pi_2(\mathbf{x}^\beta)\right\rangle\right|}}$$

If also, for distinct arcs $(i \to j)$ and $(p \to q)$, $|<M_{ij}(\mathbf{x}^\alpha)>| = |<M_{ij}(\mathbf{x}^\beta)>| = |<M_{pq}(\mathbf{x}^\alpha)>| = |<M_{pq}(\mathbf{x}^\beta)>|$ (as would be the case if the magnitudes of the arc weights were all randomly drawn from the same distribution), then

(16)
$$\frac{\left|\left\langle\Pi_1(\mathbf{x}^\alpha)\Pi_2(\mathbf{x}^\beta)\right\rangle\right|}{\sqrt{\left|\left\langle\Pi_1(\mathbf{x}^\alpha)\Pi_2(\mathbf{x}^\alpha)\right\rangle\left\langle\Pi_1(\mathbf{x}^\beta)\Pi_2(\mathbf{x}^\beta)\right\rangle\right|}}$$
$$= \frac{\left|\left\langle M_{ij}(\mathbf{x}^\alpha)\right\rangle\left\langle M_{pq}(\mathbf{x}^\beta)\right\rangle\right|}{\sqrt{\left|\left\langle M_{ij}(\mathbf{x}^\alpha)\right\rangle\left\langle M_{ij}(\mathbf{x}^\beta)\right\rangle\left\langle M_{pq}(\mathbf{x}^\alpha)\right\rangle\left\langle M_{pq}(\mathbf{x}^\beta)\right\rangle\right|}} \frac{\left|\left\langle\Pi_1(\mathbf{x}^\alpha)\Pi_2(\mathbf{x}^\beta)\right\rangle\right|}{\sqrt{\left|\left\langle\Pi_1(\mathbf{x}^\alpha)\Pi_2(\mathbf{x}^\alpha)\right\rangle\left\langle\Pi_1(\mathbf{x}^\beta)\Pi_2(\mathbf{x}^\beta)\right\rangle\right|}}$$
$$= \frac{\left|\left\langle[M_{ij}(\mathbf{x}^\alpha)\Pi_1(\mathbf{x}^\alpha)][M_{pq}(\mathbf{x}^\beta)\Pi_2(\mathbf{x}^\beta)]\right\rangle\right|}{\sqrt{\left|\left\langle[M_{ij}(\mathbf{x}^\alpha)\Pi_1(\mathbf{x}^\alpha)][M_{pq}(\mathbf{x}^\alpha)\Pi_2(\mathbf{x}^\alpha)]\right\rangle\left\langle[M_{ij}(\mathbf{x}^\beta)\Pi_1(\mathbf{x}^\beta)][M_{pq}(\mathbf{x}^\beta)\Pi_2(\mathbf{x}^\beta)]\right\rangle\right|}}$$

Finally, if it may be assumed that $<\Pi_1 \Pi_2> \gg <\Pi_1><\Pi_2>$ (with the quantities $\Pi_1$ and $\Pi_2$ evaluated at either fixed point $\mathbf{x}^\alpha$ or $\mathbf{x}^\beta$) – as might be the case for products of suitably many arc weights even if the variation in individual arc weights is small compared to their average values – then the brackets in **equations (15)-(16)** above may be replaced with double brackets, leading to

(17)
$$\frac{\left|\left\langle\left\langle[M_{ij}(\mathbf{x}^\alpha)\Pi_1(\mathbf{x}^\alpha)][M_{ij}(\mathbf{x}^\beta)\Pi_2(\mathbf{x}^\beta)]\right\rangle\right\rangle\right|}{\sqrt{\left|\left\langle\left\langle[M_{ij}(\mathbf{x}^\alpha)\Pi_1(\mathbf{x}^\alpha)][M_{ij}(\mathbf{x}^\alpha)\Pi_2(\mathbf{x}^\alpha)]\right\rangle\right\rangle\left\langle\left\langle[M_{ij}(\mathbf{x}^\beta)\Pi_1(\mathbf{x}^\beta)][M_{ij}(\mathbf{x}^\beta)\Pi_2(\mathbf{x}^\beta)]\right\rangle\right\rangle\right|}}$$
$$\leq \frac{\left|\left\langle\left\langle[M_{ij}(\mathbf{x}^\alpha)\Pi_1(\mathbf{x}^\alpha)][M_{pq}(\mathbf{x}^\beta)\Pi_2(\mathbf{x}^\beta)]\right\rangle\right\rangle\right|}{\sqrt{\left|\left\langle\left\langle[M_{ij}(\mathbf{x}^\alpha)\Pi_1(\mathbf{x}^\alpha)][M_{pq}(\mathbf{x}^\alpha)\Pi_2(\mathbf{x}^\alpha)]\right\rangle\right\rangle\left\langle\left\langle[M_{ij}(\mathbf{x}^\beta)\Pi_1(\mathbf{x}^\beta)][M_{pq}(\mathbf{x}^\beta)\Pi_2(\mathbf{x}^\beta)]\right\rangle\right\rangle\right|}}$$

The assumptions of statistical independence and equal average magnitude of the arc weightings $M_{ij}(\mathbf{x}^\alpha)$ made above cannot be expected to hold rigorously when the distributions of arc weights are derived from the fixed points of dynamical systems such as those described in **appendix 1**. Nevertheless, one might still hypothesize that **inequality (17)** could offer some insight into the "average impact" of arc overlap in the various products of arc weights described herein as the ($k,l$)-terms of a system. This conjecture is the essence of the differential overlap-dependence hypothesis (DODH) described in **statements (24)-(25)** below. (Simulation results aimed specifically at investigating the validity of **inequality (17)** for the systems described in **appendix 1** are discussed below in this section (and see **table 1**).)

Applied specifically to the ($k,l$)-terms of a system, the DODH then suggests that, all else being equal, the presence of overlap arcs in a ($k,l$)-term $A=K_A \cup L_A$ tends to decrease the ratio

(18)
$$\frac{\left|\left\langle\left\langle K_A(\mathbf{x}^\alpha)L_A(\mathbf{x}^\beta)\right\rangle\right\rangle\right|}{\sqrt{\left|\left\langle\left\langle K_A(\mathbf{x}^\alpha)L_A(\mathbf{x}^\alpha)\right\rangle\right\rangle\left\langle\left\langle K_A(\mathbf{x}^\beta)L_A(\mathbf{x}^\beta)\right\rangle\right\rangle\right|}}$$

of the magnitudes of the contributions of the ($k,l=k$)-term $A$ to the numerator and denominator, respectively, of **equation (1)** for the correlation Corr($F_k(\mathbf{x}^\alpha),F_k(\mathbf{x}^\beta)$).

The degree to which **ratio (18)** is decreased by the overlap of $K_A$ and $L_A$ at a particular arc ($i \rightarrow j$) may be expected to depend strongly on the intensity of the interactions of node $i$ and $j$ with whatever switch(es) in the system lead to the distinction of the states $\mathbf{x}^\alpha$ and $\mathbf{x}^\beta$. For example, if nodes $i$ and $j$ were totally disconnected from any switches, $M_{ij}(\mathbf{x}^\alpha)$ would necessarily be equal to $M_{ij}(\mathbf{x}^\beta)$, so that Corr($M_{ij}(\mathbf{x}^\alpha),M_{ij}(\mathbf{x}^\beta)$)=1. Then the assumptions made above would lead to the equality sign holding in **inequality (17)**, so that there would be no expected reduction whatsoever in **ratio (18)**. Thus it seems that only when nodes $i$ and/or $j$ have significant interaction with the switch driving the distinction between the different fixed points should overlap at the arc ($i \rightarrow j$) be expected to have a significant impact on **ratio (18)**, and hence on Corr($F_k(\mathbf{x}^\alpha),F_k(\mathbf{x}^\beta)$).

To test the predictions made above with regard to the effects of overlapping arcs at varying distances from a switch on the **ratio (18)**, bistable switch-containing systems were constructed and subjected to perturbations according to the method described in **appendix 3** for further analysis. Define

(19) $\Gamma_{k,l}^{(ab)(cd)}(\mathbf{x}^\alpha,\mathbf{x}^\beta) = \sum_{A \in \Theta_{k,l}} \sum_{\{K_A,L_A\} \in D_A^{(k,l)} | (a \rightarrow b) \in K_A \wedge (c \rightarrow d) \in L_A} \left\langle\left\langle K_A(\mathbf{x}^\alpha)L_A(\mathbf{x}^\beta)\right\rangle\right\rangle$

(i.e., the sum of all ($k,l$)-terms $A$ in which the $k$-term $K_A$ contains the arc ($a \rightarrow b$) and the $l$-

| Arc Set $X_\delta$ | $\Gamma^\delta_{cross:same}$ | $\Phi^\delta_{cross:same}$ |
|---|---|---|
| $X_0$ | 1.007 ± 0.024 | 0.946 ± 0.007 |
| $X_1$ | 0.829 ± 0.025 | 0.675 ± 0.031 |
| $X_2$ | 0.834 ± 0.023 | 0.795 ± 0.022 |

**Table 1.** Estimated median (±associated standard error) values of $\Gamma^\delta_{cross:same}$ (defined by **equation (22)**) and $\Phi^d_{cross:same}$ (**equation (23)**) for random digraph systems ($n=26$, $<d>=4$) constructed as described in **appendix 3** and analyzed as described in **appendix 4**. Note that the data indicate that $\Gamma^\delta_{cross:same}$ is usually larger than $\Phi^\delta_{cross:same}$, with the magnitude of this difference varying strongly with the distance $\delta$ (of the nodes used in defining $\Gamma$ and $\Phi$) away from the switch.

term $L_A$ contains the arc ($c \to d$)), and

(20) $$\Phi^{(ab)(cd)}_{k,l}\left(\mathbf{x}^\alpha, \mathbf{x}^\beta\right) = \left|\Gamma^{(ab)(ab)}_{k,l}\left(\mathbf{x}^\alpha, \mathbf{x}^\beta\right)\right| + \left|\Gamma^{(cd)(cd)}_{k,l}\left(\mathbf{x}^\alpha, \mathbf{x}^\beta\right)\right|$$

(i.e., the sum of all ($k,l$)-terms $A$ which *overlap* at either the arc ($a \to b$) or the arc ($c \to d$). Define also three distinct sets $N_0$, $N_1$, and $N_2$ of nodes distinguished by distance from the switch nodes:

(21) $N_0 = \{i \mid \text{Node } i \text{ is a switch node}\}$

$N_1 = \{i \mid \text{Node } i \text{ neighbors a switch node but is not itself a switch node}\}$

$N_2 = \{i \mid \text{Node } i \text{ neither is a switch node nor neighbors a switch node}\}$

Now, for each node set $N_\delta$, define a set $X_\delta$ of 100 randomly selected pairs of arcs adjacent to nodes of $N_\delta$ as follows. [1] Choose a node from the set $N_\delta$ (with uniform probability) and a direction (either in or out). If the chosen node does not have two distinct arcs incident upon it with the required directionality, start step [1] over. [2] Randomly choose (uniform probability) two distinct arcs adjacent to the chosen node with the required directionality and add this pair to the set $X_\delta$ (the newly added pair may be the same as an earlier pair chosen for inclusion in the $X_\delta$). Repeat until 100 (not necessarily distinct) arc pairs have been added to $X_\delta$.

**Table 1** lists estimates for of typical (median) values of the ratios of arc-set-$X_\delta$-averaged magnitudes of $\Gamma$ evaluated across different fixed points divided by the same quantities $\Gamma$ evaluated with both arguments set to the same fixed point:

(22) 
$$\Gamma^\delta_{cross:same} = \left.\frac{\sum_{\{(a\to b),(c\to d)\} \in X_\delta} \left|\Gamma^{(ab)(cd)}_{k,l}\left(\mathbf{x}^\alpha, \mathbf{x}^\beta\right)\right|}{\sqrt{\left[\sum_{\{(a\to b),(c\to d)\} \in X_\delta} \left|\Gamma^{(ab)(cd)}_{k,l}\left(\mathbf{x}^\alpha, \mathbf{x}^\alpha\right)\right|\right]\left[\sum_{\{(a\to b),(c\to d)\} \in X_\delta} \left|\Gamma^{(ab)(cd)}_{k,l}\left(\mathbf{x}^\beta, \mathbf{x}^\beta\right)\right|\right]}}\right|_{k=l=n}$$

Note that only ($k,l$)-terms which contain but do not overlap at the arcs ($a \to b$) and ($c \to d$) from the pairs in $X_\delta$ contribute to **equation (22).** These are contrasted in **table 1** with estimates of similar "cross fixed point:same fixed point" ratios of $\Phi$:

(23)
$$\Phi_{\text{cross:same}}^{\delta} = \left. \frac{\sum_{\{(a\to b),(c\to d)\}\in X_\delta} \left|\Phi_{k,l}^{(ab)(cd)}\left(\mathbf{x}^\alpha, \mathbf{x}^\beta\right)\right|}{\sqrt{\left[\sum_{\{(a\to b),(c\to d)\}\in X_\delta} \left|\Phi_{k,l}^{(ab)(cd)}\left(\mathbf{x}^\alpha, \mathbf{x}^\alpha\right)\right|\right]\left[\sum_{\{(a\to b),(c\to d)\}\in X_\delta} \left|\Phi_{k,l}^{(ab)(cd)}\left(\mathbf{x}^\beta, \mathbf{x}^\beta\right)\right|\right]}} \right|_{k=l=n}$$

to which only those ($k,l$)-terms which overlap at arcs from the pairs in $X_\delta$ contribute. **Appendix 4** describes the methods used in obtaining the data shown in **table 1**.

The data presented in **table 1** support the contention that the **ratio (18)** of the contribution of a given ($k,k$)-term $A$ to the numerator divided by the contribution of $A$ to the denominator of **equation (1)** for $\text{Corr}(F_k(\mathbf{x}^\alpha),F_k(\mathbf{x}^\beta))$ tends to be smaller when $A$ overlaps at a particular arc than when it does not – at least, when the termini of the arc in question lie sufficiently nearby the relevant switch(es) controlling the distinction between $\mathbf{x}^\alpha$ and $\mathbf{x}^\beta$. It is clear from comparing the data for the arc-pair sets $X_0$, $X_1$, and $X_2$ that the impact of arc overlap is highly dependent on the distance of the nodes involved from the switch nodes. It is interesting to note that the effects of overlap appear to be maximized for arcs incident upon the neighbors of the switch nodes, as opposed to the switch nodes themselves (see **appendix 3** for a brief discussion of this phenomenon). Thus, the term "relevant arcs" is here introduced to describe describe such arcs in developing the DODH below.

It should be noted that **equations (22)-(23)** restrict the values of $k$ and $l$ to the maximum possible value $n$. This was done because, as is evident from the data in **table 1**, the distance of the arc nodes contained within a ($k,l$)-term from any relevant switch nodes strongly influences the values of **equations (22)-(23)**. However, any ($n,n$)-term $A$ will contain arcs entering and leaving *each* node of the system exactly once in both the $k$-term $K_A$ and the $l$-term $L_A$, so that the distribution of arc nodes is for such ($n,n$)-terms always the same.

It is now useful to offer the first statement of the DODH:

(24) *Differential Overlap Dependence Hypothesis* 1: Network topologies in which ($k,l$)-terms overlap more frequently at relevant arcs will tend to produce lower values of $\text{Corr}(F_k(\mathbf{x}^\alpha),F_k(\mathbf{x}^\beta))$ than topologies with less relevant ($k,l$)-term overlap.

This form of the DODH may be applied to understand the impact of network topology on $\text{Corr}(F_k(\mathbf{x}^\alpha),F_k(\mathbf{x}^\beta))$ as long as the relationship between network topology and relevant arc overlap can be established. Note that as the in-degree of a node increases, the number of ways to choose two distinct arcs entering said node increases more quickly than does the number of ways to choose a single arc entering the node (with a similar statement applying to the out-degree and arcs leaving a node). That is, increasing the degree of a node should be generally expected to decrease the fraction of ($k,l$)-terms which overlap at arcs entering or leaving the node in question. Combined with the observation that the degrees of nodes neighboring switch nodes tend to increase with both the average and variance of the network degree distribution, this suggests that relevant arc overlap is decreased in denser and/or more variable-degree networks.

**Statement (24)** of the differential overlap dependence hypothesis may thus be compared with the results for $\text{Corr}(F_k(\mathbf{x}^\alpha),F_k(\mathbf{x}^\beta))$ (see **figure 5**) obtained in simulations of systems again constructed and perturbed by the method described in **appendix 3**. The correlation of the $F_k$'s for the two different switch settings is indeed observed to be higher for both the denser and higher degree-variance networks.

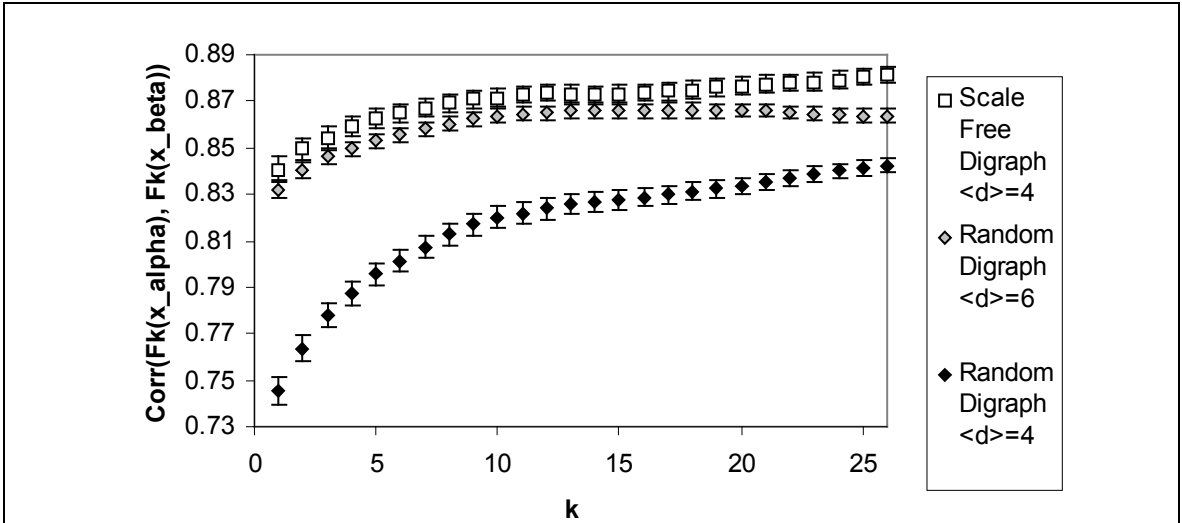

**Figure 5.** Estimated median value of Corr($F_k(\mathbf{x}^\alpha),F_k(\mathbf{x}^\beta)$) as a function of $k$ [$n$=26] for systems constructed as indicated in **appendix 5**. Error bars indicate standard error of median estimates.

Finally, as discussed at the end of **section 2**, it is here conjectured that Corr($F_k(\mathbf{x}^\alpha),F_k(\mathbf{x}^\beta)$) should be inversely related to the likelihood of switch integration. This yields the second form of the DODH:

(25) *Differential Overlap Dependence Hypothesis* 2: Switch integration is more likely in networks with topologies with more relevant ($k,l$)-term overlap – e.g., networks with lower average or variance of the degree distribution – than in networks with topologies with less relevant ($k,l$)-term overlap.

The results shown in **figure 2** of the simulations described in **appendix 1** appear to be consistent with this form of the DODH.

## 5. Discussion and Conclusions

It is interesting that the results of this study indicate that topologies characterized by highly skewed degree distributions lead to low propensity for switch integration, especially in light of the finding that many biological networks one might expect to be "integrating switches" have been characterized as approximately scale-free in structure [3-5, 8-10, 47]. Various suggestions have been advanced to explain the appearance of this sort of structure, stressing both possible advantageous properties, such as various sorts of robustness under node removal [9, 32, 48], and biological mechanisms (e.g., gene duplication) which would tend to form scale-free structures [32, 49-51]. On the other hand, Amaral et. al. [47] focus on the existence of exponential cutoffs of the degree distributions of many networks which appear to follow a power-law below the cutoff. In particular, they show that constraints which limit the addition of new arcs to vertices that already have many can naturally produce such patterns. The results presented here might then suggest switch integration may pose one such constraint for some networks.

Alternatively, switch-integrating networks may be structured in such a way that those parts of the network which are most focused on the task of switch integration are described by degree distributions with (relatively) small variance compared to that of the network as a whole. Thus one might envision searching for potential switch-integrating subnetworks by looking for sets of nodes of relatively homogenous (low) degree strongly linked to each other by feedback loops not passing through network hubs. This attractive scenario suggests further study of switch integration in networks with more complicated structure than those considered here. Such extensions of the results reported herein to consideration of more general network structures could also provide further insight into what to look for to identify real biological networks and subnetworks which might integrate their switches.

The finding here that switch integration becomes less likely as arc density increases may offer yet another interesting wrinkle to the ongoing stability-complexity discussion in the ecology literature [20-23]. If some of the variation in overall community structure resulting from the perturbation/removal of one or a few species results from processes similar to the switch integration phenomenon discussed here, then it would seem that increasing the "complexity" of an ecosystem by increasing its interaction density might have some tendency to increase its robustness. That is, toggling the state of one "switch" by (say) removing a species which participates in it would be less likely to result in disturbing the community structure by shifting the states of other switches in more densely interconnected networks than in sparser ones. Similarly, this line of thinking would suggest that increased variance of degree distribution might act to increase the robustness of a community by depressing switch integration.

On the other hand, there may be some situations in which an ecological community benefits from the ability to integrate switches. If such a community is exposed to periodically varying environmental conditions throughout its history, it is likely that different competitors will thrive at different times. In this case, communities in which such competitive switches act in concert to achieve a community-wide transition might undergo less stress in the transient periods than those in which the switches work independently. Over time, those constituent parts of a community network which achieve such an integrated response might thus retain their structure more faithfully than those parts of the network which do not, ultimately leading to an increase in integration-promoting structure. Of course, if suddenly subjected to a new sort of disturbance unlike those to which the community has historically been subjected, those parts of the community with less integration-promoting structures might prove more robust, as discussed in the paragraph above.

With regard to ecological applications of this switch integration theory, it should be noted that the network models studied here did not include any trophic structure. It would be of great interest in future studies extending the switch integration approach to more complicated and/or general types of networks to explicitly consider how trophic stratification shapes the relevant structures.

It should be stressed that several key mathematical conjectures were made in arriving at the conclusions of this study, especially the "differential overlap dependence hypothesis" described in **section 4**, with support provided by recourse to computer simulations. The author suspects that there are some interesting lessons to be learned in further attempts to appropriately qualify and verify these conjectures.

**Appendix 1. Computer Simulations of Switch-Containing Random Dynamical Systems**

The random dynamical systems used here were generated by: (1) generating a random digraph or scale-free digraph as described in **appendix 2**, and (2) adding reactions in accord with the topology thus defined (with the exception that some arcs will be added regardless of their presence in this pre-defined topology as part of the process of embedding the bistable switches).

The two-dimensional dynamical system defined by

(26)
$$\frac{dx_1}{dt} = \frac{1}{4} + x_1 - \frac{(x_1)^2}{5} - x_1 x_2$$
$$\frac{dx_2}{dt} = \frac{1}{4} + x_2 - \frac{(x_2)^2}{5} - x_1 x_2$$

exhibits bistability, with stable fixed points at (0.0633, 4.9367) and (4.9367, 0.0633). Four of the nodes of each random digraph were associated with two copies of this system, so that these four nodes are subdivided into two sets of two nodes each, with arcs going both ways connecting the two nodes within each such set. Again, these arcs were added regardless of their presence or absence in the pre-defined randomly generated topology. Topologies in which there was not a path connecting each of the two two-node switches to the other were excluded from further consideration.

For each of the remaining nodes of the system, reactions associated with one-loop arcs ($i{\rightarrow}i$) of the form

(27)
$$\frac{dx_i}{dt} = a_i - b_i x_i + \ldots$$

were added. The parameters $a_i$ and $b_i$ were chosen from a log-normal distribution, with $<\ln(a_i)> = <\ln(b_i)> = \ln(0.1)$, and $<<\ln(a_i)^2>> = <<\ln(b_i)^2>> = 0.2$.

For each remaining arc ($i{\rightarrow}j$) in the system, one of four types of reaction was added, with the type chosen with uniform probability from the set $\{1,2,3,4\}$. It should be noted that each of these reaction types required the specification of exactly one rate constant $c_{ji}$; in all cases, this parameter was chosen from a log-normal distribution with $<\ln(c_{ji})> = \ln(0.075)$, $<<\ln(c_{ji})^2>> = 1$ (with $c_{ji}$ independent of $c_{lk}$ unless $j = l$ and $i = k$).

(28) Type 1: (species $i \rightarrow$ species $j$)
$$\frac{dx_i}{dt} = \ldots - c_{ji} x_i + \ldots$$
$$\frac{dx_j}{dt} = \ldots + c_{ji} x_i + \ldots$$

(29) Type 2: (species $i \rightarrow$ species $i$ +species $j$)
$$\frac{dx_i}{dt} = \ldots$$
$$\frac{dx_j}{dt} = \ldots + c_{ji} x_i + \ldots$$

(30) Type 3: (species $i$ + species $j \rightarrow$ species $i$)

$$\frac{dx_i}{dt} = \ldots$$

$$\frac{dx_j}{dt} = \ldots - c_{ji} x_i x_j + \ldots$$

(31) Type 4: (species $i$ + species $i$ → species $i$ + species $j$)

$$\frac{dx_i}{dt} = \ldots - c_{ji}(x_i)^2 + \ldots$$

$$\frac{dx_j}{dt} = \ldots + c_{ji}(x_i)^2 + \ldots$$

Each such randomly generated dynamical system was then tested for switch integration. The first step in this procedure was to start the system successively at each of the four points in phase space described by (noting that $x_1$ and $x_2$ are taken to be the components of the first embedded switch, while $x_3$ and $x_4$ are taken as the components of the second embedded switch):

(32)
$$\begin{pmatrix} x_1 = 4.9367 \\ x_2 = 0.0633 \\ x_3 = 4.9367 \\ x_4 = 0.0633 \\ x_i = 0, \ i > 4 \end{pmatrix}, \begin{pmatrix} x_1 = 0.0633 \\ x_2 = 4.9367 \\ x_3 = 4.9367 \\ x_4 = 0.0633 \\ x_i = 0, \ i > 4 \end{pmatrix}, \begin{pmatrix} x_1 = 4.9367 \\ x_2 = 0.0633 \\ x_3 = 0.0633 \\ x_4 = 4.9367 \\ x_i = 0, \ i > 4 \end{pmatrix}, \begin{pmatrix} x_1 = 0.0633 \\ x_2 = 4.9367 \\ x_3 = 0.0633 \\ x_4 = 4.9637 \\ x_i = 0, \ i > 4 \end{pmatrix}$$

(these can be thought of as, e.g., states (on,on), (off,on), (on,off), and (off,off) with regard to the two embedded switch systems when they are removed from the surrounding system) and then numerically computing their evolution for 100 time units using the MATLAB routine ode15s. At the end of each of these trajectories, the nearest fixed point of the dynamics was located with MATLAB routine fsolve.

If a system exhibited exactly two distinct stable fixed points as a result of this procedure, *and* if it was true of each of these stable fixed points (*sfps*) $\mathbf{x}^\alpha$ satisfied one of the four (mutually exclusive) "switch-conditions"

(33)  I:  $x_1^\alpha > 10 x_2^\alpha$,  $x_3^\alpha > 10 x_4^\alpha$

II:  $10 x_1^\alpha < x_2^\alpha$,  $x_3^\alpha > 10 x_4^\alpha$

III:  $x_1^\alpha > 10 x_2^\alpha$,  $10 x_3^\alpha < x_4^\alpha$

IV:  $10 x_1^\alpha < x_2^\alpha$,  $10 x_3^\alpha < x_4^\alpha$

(these conditions require the "on" and "off" switch states in the full system to qualitatively resemble the "on" and "off" states in the isolated switch systems), then the system was examined further. In this case, the number of "switch-flips" exhibited by the system was defined as follows:

0, if both sfps satisfy the same condition from **expression (33)**.
1, if one sfp satisfies I or IV and the other II or III.
2, if one sfp satisfies I and the other IV, or if one sfp satisfies II and the other III.

Systems with zero switch-flips (i.e., both stable fixed points satisfying the same switch-condition, **expression (33)**) were excluded from further consideration, as they did not share the qualitative behavior of the embedded switches (**equation (26)**) isolated from the full system.

The propensity of different topologies toward switch integration was then finally measured by considering the ratio of the count of those systems that exhibited two switch-flips to the count of those that exhibited either one or two switch-flips. That is, those systems with two switch-flips were regarded as exhibiting switch integration, while those with only one were regarded as exhibiting one non-integrated switch. The results, displayed as the fraction of observed switches thus defined which exhibited integration, are shown in **figure 2**.

**Appendix 2. Generation of Digraph Topologies**

Random digraph topologies of *n* nodes of average in-/out-degree <*d*> were generated by including each (non-one loop) arc ($i \rightarrow j$) with independent probability (<*d*>/(*n*-1)). All one loops ($i \rightarrow i$) were included in every random digraph used herein (but were here *not* counted in the in- or out-degree of the nodes).

Scale-free digraph topologies (with, in this paper, average node in- and out-degree always equal to four) were generated via a preferential attachment mechanism, similar to that of Barabasi, et. al. [36]. First, a set of five fully connected nodes was generated. An iterated procedure in which nodes were added one-by-one, with four new arcs added for each new node, was then followed until the desired system size was reached. For each new node *i*, two arcs were added each with their tail in *i* and their head chosen to be in previously added node $j < i$ with probability

(34) $$p^{(i)}_{ji} = \frac{d^{in}_j + d^{out}_j}{\sum_{k=1}^{(i-1)} \left(d^{in}_k + d^{out}_k\right)}$$

where $d^{in}_j$ [$d^{out}_j$] is the in-[out-]degree of node *j* before the arc in question is added to the system. Similarly, two arcs were added each with their head in the new node *i* and their tail chosen to be in previously added node *j* with probability $p^{(i)}_{ij} = p^{(i)}_{ji}$, given by **equation (34)**.

This procedure generates topologies which, for $n \rightarrow \infty$, satisfy [36]

(35) $$\left\langle\left\langle\left(d^{out}_i\right)^2\right\rangle\right\rangle = \left\langle\left\langle\left(d^{in}_i\right)^2\right\rangle\right\rangle \propto \ln n$$

For the switch integration simulations (**appendix 1**) performed with scale-free topologies constructed in this manner, the nodes were shuffled randomly before assigning the lowest numbered nodes to be associated with the embedded switches.

**Appendix 3. Computational Tests of Differential Overlap Dependence Hypothesis**

To test different aspects of the differential overlap dependence hypothesis (specifically **inequality (17)** and **statement (24)**), simulations were performed on dynamical systems with varying topologies generated as in **appendices 1-2**, with two modifications. First, only one copy of the bistable switch **equation (26)** was embedded into these systems, associated with nodes 1 and 2. Only systems which remained bistable after the coupling of the switch to the network, and for which the two stable fixed points $\mathbf{x}^1$ and $\mathbf{x}^2$ satisfied (when ordered correctly)

(36)  $10x^1_1 < x^1_2$ and $x^2_1 > 10x^2_2$

(analogous to **inequalities (33)** in **appendix 1**, but with only one switch) were considered. The second modification to the method of constructing systems described in **appendix 1** was the further requirement that

(37) $\quad$ either $15x_1^1 > x_2^1$ or $x_1^2 < 15x_2^2$

The requirements of **inequalities (36)-(37)** were imposed because the strength of the interaction of the switch with the network tends to vary according to both the magnitude of the individual arc weightings and the density of the arcs in the system network. **Inequalities (36)-(37)** control for this differential size-of-perturbation effect in comparing different network structures. Thus, only systems which still contained a switch sufficiently similar to the decoupled switch (**inequality (36)**), but for which at least one setting of the coupled switch was sufficiently destabilized (**inequality (37)**), were compared. (The requirement that at least one switch setting be sufficiently destabilized can be thought of as analogous to demanding that the switch should be susceptible to having at least one setting made unavailable by total destabilization if the system were to be perturbed by coupling to a second switch.)

Next, in order to get an estimate for Corr($F_k(\mathbf{x}^\alpha),F_k(\mathbf{x}^\beta)$), 100 different sets of values for the parameters $\{a_i\}$ were generated, with, in each case, each $a_i$ chosen from a log-normal distribution with $<\ln(a_i)>=\ln(a_i^0)$ (where $a_i^0$ is the unperturbed value of $a_i$ for the system in question) and $<<\ln(a_i)^2>>=0.05$. These slight variations of the parameters $\{a_i\}$ represent a set of distinct perturbations to the system. The two fixed points of each of the 100 resulting "perturbed" systems were located and the characteristic polynomial of the matrix of the linearized dynamics calculated at each fixed point (using the MATLAB routine poly). This data was then used to estimate Corr($F_k(\mathbf{x}^\alpha),F_k(\mathbf{x}^\beta)$) for the system.

(Note that none of the parameters directly associated with the switch nodes were varied in the procedure just described. Thus, the only source of perturbation in the fixed point values for the switch node variables was the coupling to the remainder of the network. This likely underlies the phenomenon (shown in **table 1**) of arc overlap having maximal impact for nodes neighboring switch nodes, as opposed to arcs associated directly with switch nodes. This may be compared to the "switch integration" setup described in **appendix 1**, for which the perturbation of the states of one switch result from its coupling through the network to the states of another switch.)

The procedure described above was repeated for 2,500 different random digraph topologies with $<d>=4$ (divided into 25 groups of 100 each), 2,500 (25 groups of 100) random digraph topologies with $<d>=6$, and 2,500 (again, 25 groups of 100) different scale-free ($<d>=4$) topologies generated as described in **appendix 2**. The median values of Corr($F_k(\mathbf{x}^\alpha),F_k(\mathbf{x}^\beta)$) of each group of 100 were averaged to obtain the results displayed in **figure 5**. Medians were used because the variation in the mean values of the different groups was significantly larger than that of the medians, suggesting a highly skewed distribution.

**Appendix 4. Isolation of Sums of *k*-terms Containing a Particular Arc**

The sum of all *k*-terms present in a (real) matrix $M=M(\mathbf{x}^\alpha)$ passing through a particular arc ($a \rightarrow b$) may be calculated by considering a related matrix $\hat{M}^{(ab)}(\mathbf{x}^\alpha)$ defined as follows:

(38) $$\hat{M}_{pq}^{(ab)}(\mathbf{x}^\alpha) = \begin{cases} iM_{pq}(\mathbf{x}^\alpha), & \text{if } p = a \text{ and } q = b \\ M_{pq}(\mathbf{x}^\alpha), & \text{otherwise} \end{cases}.$$

(That is, the arc $(a \to b)$ is "labeled" by multiplication by the imaginary number $i$.) Then all $k$-terms in $\hat{M}^{(ab)}(\mathbf{x}^\alpha)$ containing the arc $(a \to b)$ will be purely imaginary, while all $k$-terms not containing $(a \to b)$ will be purely real, so that (defining $\hat{F}_k^{(ab)}(\mathbf{x}^\alpha)$ to be the coefficients of the characteristic polynomial of the matrix $\hat{M}^{(ab)}(\mathbf{x}^\alpha)$)

(39) $$\text{Im}\left[\hat{F}_k^{(ab)}(\mathbf{x}^\alpha)\right] = \sum_{\{K \in \Theta_k | (a \to b) \in K\}} K(\mathbf{x}^\alpha)$$

Now consider the $(k,l)$-terms in $\hat{F}_k^{(ab)}(\mathbf{x}^\alpha)\hat{F}_l^{(cd)}(\mathbf{x}^\beta)$. Those for which the $k$-term contains the arc $(a \to b)$ or the $l$-term contains the arc $(c \to d)$ – but not both at once – will be purely imaginary, so that

(40) $$\text{Im}\left[\left\langle\left\langle \hat{F}_k^{(ab)}(\mathbf{x}^\alpha)\hat{F}_l^{(cd)}(\mathbf{x}^\beta)\right\rangle\right\rangle\right] = \sum_{A \in \Theta_{k,l}} \sum_{\{K_A, L_A\} \in D_A^{(k,l)} | (a \to b) \in K_A \,\dot\vee\, (c \to d) \in L_A} \left\langle\left\langle K_A(\mathbf{x}^\alpha)L_A(\mathbf{x}^\beta)\right\rangle\right\rangle$$

(where $\dot\vee$ in **equation (40)** represents the exclusive or (XOR) operator).

Those $(k,l)$-terms which contain $(a \to b)$ in the $k$-term and simultaneously $(c \to d)$ in the $l$-term, on the other hand, will be purely real, but with reversed sign in $\hat{F}_k^{(ab)}(\mathbf{x}^\alpha)\hat{F}_l^{(cd)}(\mathbf{x}^\beta)$ as compared with $F_k(\mathbf{x}^\alpha)F_l(\mathbf{x}^\beta)$ resulting from the product of the two factors of $i$. Those remaining $(k,l)$-terms $A$, for which $(a \to b)$ is not part of $K_A$ and $(c \to d)$ is not part of $L_A$, remain real with unchanged sign. Thus,

(41) $$\text{Re}\left[\left\langle\left\langle F_k(\mathbf{x}^\alpha)F_l(\mathbf{x}^\beta)\right\rangle\right\rangle\right] - \text{Re}\left[\left\langle\left\langle \hat{F}_k^{(ab)}(\mathbf{x}^\alpha)\hat{F}_l^{(cd)}(\mathbf{x}^\beta)\right\rangle\right\rangle\right] - \text{Im}\left[\left\langle\left\langle \hat{F}_k^{(ab)}(\mathbf{x}^\alpha)\hat{F}_l^{(cd)}(\mathbf{x}^\beta)\right\rangle\right\rangle\right]$$

$$= 2 * \sum_{A \in \Theta_{k,l}} \sum_{\{K_A, L_A\} \in D_A^{(k,l)} | (a \to b) \in K_A \,\wedge\, (c \to d) \in L_A} \left\langle\left\langle K_A(\mathbf{x}^\alpha)L_A(\mathbf{x}^\beta)\right\rangle\right\rangle$$

$$= 2\Gamma_{k,l}^{(ab)(cd)}(\mathbf{x}^\alpha, \mathbf{x}^\beta)$$

This technique was used to isolate the sums of all $(k,l)$-terms containing given arc pairs, as described in **section 4**. The data in **table 1** are estimated by considering median values of 100 simulations, averaged over 50 such groups of simulations. As in **appendix 3**, median values were used because the variation in the mean values of the different groups was significantly larger than that of the medians.

**References**

1. Murray, J.D., *Mathematical Biology*. Vol. 1. 2002, New York, NY: Springer.
2. Strogatz, S.H., *Nonlinear Dynamics and Chaos*. 1994, Cambridge, MA: Perseus Books Group.
3. Jeong, H., et al., *The Large-Scale Organization of Metabolic Networks*. Nature, 2000. **407**(6804): p. 651-654.
4. Dunne, J.A., R.J. Williams, and N.D. Martinez, *Food-web Structure and Network Theory: The Role of Connectance and Size*. Proceedings of the National Academy of Sciences, 2002. **99**(20): p. 12917-12922.
5. Provero, P., *Gene Networks from DNA Microarray Data: Centrality and Lethality*. arXiv:cond-mat/0207345v2, 2002.
6. Sporns, O. and R. Kotter, *Motifs in Brain Networks*. PLoS Biology, 2004. **2**(11): p. 1910-1918.
7. Milo, R., et al., *Network Motifs: Simple Building Blocks of Complex Networks*. Science, 2002. **298**: p. 824-828.
8. Wagner, A., *The Yeast Protein Interaction Network Evolves Rapidly and Contains Few Redundant Duplicate Genes*. Molecular Biology and Evolution, 2001. **18**(7): p. 1283-1292.
9. Featherstone, D.E. and K. Broadie, *Wrestling with Pleiotropy: Genomic and Topological Analysis of the Yeast Gene Expression Network*. BioEssays, 2002. **24**: p. 267-274.
10. Giot, L., et al., *A Protein Interaction Map of Drosophila Melanogaster*. Science, 2003. **302**: p. 1727-1736.
11. Kauffman, S.A., *Control Circuits for Determination and Transdetermination*. Science, 1973. **181**(4097): p. 310-318.
12. Thomas, R. and M. Kaufman, *Multistationarity, the Basis of Cell Differentiation and Memory I: Structural Conditions of Multistationarity*. Chaos, 2001. **11**(1): p. 170-179.
13. Forgacs, G. and S.A. Newman, *Biological Physics of the Developing Embryo*. 2005, Cambridge: Cambridge University Press.
14. Laurent, M. and N. Kellershohn, *Multistability: A Major Means of Differentiation and Evolution in Biological Systems*. Trends in Biochemical Sciences, 1999. **24**(11): p. 418-422.
15. Levine, M. and E.H. Davidson, *Gene Regulatory Networks for Development*. Proceedings of the National Academy of Sciences, 2005. **102**(14): p. 4936-4942.
16. Davidson, E.H., *The Regulatory Genome: Gene Regulatory Networks in Development and in Evolution*. 2006, Burlington, MA: Academic Press.
17. Crossman, A.R. and D. Neary, *Neuroanatomy: An Illustrated Colour Text*. 3 ed. 2006, Ediburgh: Churchill Livingstone.
18. Morin, P.J., *Community Ecology*. 1999, Malden, MA: Blackwell Science.
19. Sole, R.V. and J. Bascompte, *Self-Organization in Complex Ecosystems*. 2006, Princeton, NJ: Princeton University Press.
20. MacArthur, R., *Fluctuations of Animal Populations and a Measure of Community Stability*. Ecology, 1955. **36**(3): p. 533-536.
21. May, R.M., *Stability and Complexity in Model Ecosystems*. 1973, Princeton, NJ: Princeton University Press.
22. McCann, K.S., *The Diversity-Stability Debate*. Nature, 2000. **405**: p. 228-233.
23. Pimm, S.L., *The Complexity and Stability of Ecosystems*. Nature, 1984. **307**: p. 321-326.
24. Tyson, J.J. and H.G. Othmer, *The Dynamics of Feedback Control Circuits in Biochemical Pathways*. Progress in Theoretical Biology, 1978. **5**: p. 1-62.
25. Othmer, H.G., *The Qualitative Dynamics of a Class of Biochemical Control Circuits*. The Journal of Mathematical Biology, 1976. **3**: p. 53-78.
26. Logofet, D.O., *Matrices and Graphs: Stability Problems in Mathematical Ecology*. 1993, Boca Raton, FL: CRC Press.
27. Eisenfeld, J. and C. deLisi, *On Conditions for Qualitative Instability of Regulatory Circuits With Applications to Immunological Control Loops*, in *Mathematics and Computers in Biomedical Applications*, J. Eisenfeld and C. deLisi, Editors. 1985, Elsevier: Amsterdam. p. 39-53.
28. Soule, C., *Graphic Requirements for Multistationarity*. Complexus, 2003. **1**: p. 123-133.
29. Bhalla, U.S. and R. Iyengar, *Emergent Properties of Networks of Biological Signaling Pathways*. Science, 1999. **283**: p. 381-387.

30. Quirk, J. and R. Ruppert, *Qualitative Economics and the Stability of Equilibrium.* The Review of Economic Studies, 1965. **32**(4): p. 311-326.
31. Newman, M., A.-L. Barabasi, and D.J. Watts, *The Structure and Dynamics of Networks*. 2006, Princeton, NJ: Princeton University Press.
32. Barabasi, A.-L. and Z.N. Oltvai, *Network Biology: Understanding the Cell's Functional Organization.* Nature Reviews Genetics, 2004. **5**: p. 101-113.
33. Albert, R. and A.-L. Barabasi, *Statistical Mechanics of Complex Networks.* Reviews of Modern Physics, 2002. **74**: p. 47-97.
34. Aittokallio, T. and B. Schwikowski, *Graph-based Methods for Analysing Networks in Cell Biology.* Briefings in Bioinformatics, 2006. **7**(3): p. 243-255.
35. Watts, D.J. and S.H. Strogatz, *Collective Dynamics of Small-World Networks.* Nature, 1998. **393**(6684): p. 440-442.
36. Barabasi, A.-L. and R. Albert, *Emergence of Scaling in Random Networks.* Science, 1999. **286**: p. 509-512.
37. Boccaletti, S., et al., *Complex Networks: Structure and Dynamics.* Physics Reports, 2006. **424**: p. 175-308.
38. Barahona, M. and L.M. Pecora, *Synchronization in Small-World Systems.* Physical Review Letters, 2002. **89**(5): p. 054101-1-054101-4.
39. Nishikawa, T., et al., *Heterogeneity in Oscillator Networks: Are Smaller Worlds Easier to Synchronize?* Physical Review Letters, 2003. **91**(1): p. 014101-1-014101-4.
40. Hong, H., et al., *Factors That Predict Better Synchronizability on Complex Networks.* Physical Review E, 2004. **69**(6): p. 067105-1-067105-4.
41. Puccia, C.J. and R. Levins, *Qualitative Modeling of Complex Systems: An Introduction to Loop Analysis and Time Averaging*. 1986, Cambridge, MA: Harvard University Press.
42. Meinsma, G., *Elementary Proof of the Routh-Hurwitz Test.* Systems and Control Letters, 1995. **25**(4): p. 237-242.
43. Sontag, E.D., *Mathematical Control Theory: Deterministic Finite Dimensional Systems*. 1998, New York, NY: Springer.
44. Meyers, L.A., *Contact Network Epidemiology: Bond Percolation Applied to Infectious Disease Prediction and Control.* Bulletin of the American Mathematical Society, 2007. **44**(1): p. 63-86.
45. Bang-Jensen, J. and G. Gutin, *Digraphs: Theory, Algorithms and Applications*. 2002, London: Springer.
46. Landin, J., *Introduction to Algebraic Structures*. 1989, Mineola, NY: Dover.
47. Amaral, L.A.N., et al., *Classes of Small-World Networks.* Proceedings of the National Academy of Sciences, 2000. **97**(21): p. 11149-11152.
48. Sole, R.V. and J.M. Montoya, *Complexity and Fragility in Ecological Networks.* Proceedings of the Royal Society B, 2001. **268**: p. 2039-2045.
49. Sole, R.V., et al., *A Model of Large-Scale Proteome Evolution.* Advances in Complex Systems, 2002. **5**(1): p. 43-54.
50. Chung, F., et al., *Duplication Models for Biological Networks.* Journal of Computational Biology, 2003. **10**(5): p. 677-687.
51. Vazquez, A., et al., *Modeling of Protein Interaction Networks.* Complexus, 2003. **1**: p. 38-44.